\documentclass[12pt]{article}
\begin{document}
 
\title{The investigations of anisotropy in orientations of galaxies
\footnote {Presented at The Sixth Scientific Conference "Selected Issues 
of Astronomy  and Astrophysics" in honor of Bohdan Babiy 4-6 October 2011 Lviv.}
}
\medskip
 
\author{
W{\l}odzimierz God{\l}owski${^1}$ Elena Panko${^2}$\\
Paulina Pajowska ${^3}$  Piotr Flin $^{3}$
}
\maketitle
 
1. Uniwersytet Opolski, Institute of Physics, ul.  Oleska  48,
45-052 Opole, Poland e-mail:  godlowski@uni.opole.pl

2. Kalinenkov Astronomical Observatory, Nikolaev State
University, Nikolaev, Ukraine email: panko.elena@gmail.com
 
3. Uniwersytet Opolski, Institute of Physics, ul.  Oleska  48,
45-052 Opole, Poland e-mail:  paoletta@interia.pl
 
4. Pedagogical University, Institute of Physics, 25-406 Kielce, ul.
Swietokrzyska 15, Poland  e-mail: sfflin@cyf-kr.edu.pl

\section*{Abstract}
\medskip
In 1994 Parnovsky, Karachentsev and Karachentseva suggested a modified  method for
investigation of the orientations of galaxies. Using this method
they analyzed galaxies from the UGC and ESO catalogues, as well as
from their's own catalogue inclusive of flat, edge-on galaxies. They found statistically
significant anisotropy in the galaxies orientations'. In 1995 Flin
suggested that this anisotropy has to be specific to LOcal Supercluster (LSC)
In the present paper, using the method proposed by Parnovsky, Karachentsev and Karachentseva 
in 1994, we analyzed orientation of galaxies in the sample of galaxies
belonging to LSC founding only a weak anisotropy. The relation of this  method to 
Hawley and Peebles (1975) method of the investigation of the orientation of galaxies was 
discussed as well.
 
{\bf keywords}
 angular momenta, orientations of  galaxies, PACS 98.65.-r, 98.62.Ai
 
\section{Introduction}

The analysis of the anisotropy in galaxy orientation has a long history.
Usually two main methods for study of the galaxy orientations' have been introduced.
First one is based on the analysis of the galaxtic image major
axis i.e. analysis of the galactic position angles was proposed by
Hawley and Peebles \cite{h4}. Another method based on idea of
{\"O}epik \cite{Op70}, investigates the de-projection of the galaxy images,
taking into account also the galaxy's inclination with respect
to the observer's line of sight $i$. This method was applied by
Jaaniste \& Saar \cite{Ja78} and significantly modified
by Flin \& God{\l}owski \cite{f4,g2,g3,g10} (see also \cite{AR00}).
 
In 1994 Parnovsky, Karachentsev and Karachentseva \cite{Pa94} suggested
a new method for investigating the distribution of the orientations
of galaxies. This method was in fact simply a modification of the Hawley's and
Peebles's \cite{h4} approach. Parnovsky, Karachentsev and Karachentseva 
processed the data from the UGC \cite{n1} and ESO \cite{l1} catalogues,
and from their own catalogue inclusive of flat edge-on galaxies (FGC) \cite{k93}.
They detected statistically significant anisotropy in the orientation
of galaxies. They described the orientation distribution by a 
three-axis ellipsoid, showing an excess of about 20\% in the direction
$4^h-6^h$, $20^o-40^o$, and a deficit of about 25 per cent in the
direction $13^h-15^h$, $30^o- 40^o$ \cite{Pa94}. The authors recognized  that this
observed anisotropy has a global character, at least in the sense of the
survey extend. This  was generally consistent with the earlier results of
Fliche and Soriau \cite{Fl90} concerning the orientation of extended
galactic HI  envelopes and the "cosmic pole" detected in the  analysis of
remote  quasars  ($5^h  30^m, \, 7^o$). Later,   Flin  \cite{f1} analyzed
theirs results and argued that these results indicated that the observed
anisotropy is  consistent  with the results obtained previously by Flin and God{\l}owski \cite{f4}
and hasn't got a global character, instead was connected with the alignment of
galaxies in LSC.
 
The method of Parnovsky, Karachentsev and Karachentseva \cite{Pa94} was rather 
neglected in the future papers with one exception of the Parnovsky et al. analysis 
of the orientation of galaxy pairs \cite{Pa97}. In the present paper we decided  
to use the method proposed in \cite{Pa94} for analysing of the sample of galaxies 
belonging to the LSC, founding only weak possible alignment.The aim of our paper is 
repeated the Parnovsky, Karachentsev and Karachentseva \cite{Pa94} investigation using 
theirs method  on the samples of  galaxies with certain membership of Local Supercluster.
 
\section{Observational data}
 
In our work we studied the 2D-projected alignment of the galaxies in the LSC.
A sample of galaxies was taken from Tully's Nearby Galaxies (NBG) Catalogue \cite{t3}.
This Catalogue contains 2367 galaxies with radial velocities less than 
$3000\,km\,s^{-1}$. Tully's Catalogue provides relatively uniform coverage of 
the entire unobscured sky \cite{t2}. Galaxies position angles were taken 
from \cite{n1,n2,l1,l2}, while some missing measurements were made on PSS prints 
by Piotr Flin \cite{g10}. The galaxies distances were very well and in a uniform 
maner determined. As a result, the lists of galaxies were free from the background
objects. It is crucial in such type of the analysis.
 
\section{Methods of the investigations}
 
The idea of the Parnovsky, Karachentsev and Karachentseva \cite{Pa94} method 
is the analysis of the distribution
of the position angles in different coordinate systems. They assumed 
 the equatorial  coordinate system as a fundamental coordinate system 
and later the position of the coordinate system pole was varied, 
both in $\alpha$ and $\delta$. Our interest is to find the position
angles $p'$ in a particular coordinate system with respect to a pole at 
($\alpha_p, \delta_p$). The Parnovsky Karachentsev and Karachentseva function $F$ is
defined as following \cite{Pa94}:
 
\begin{equation}
F(\alpha,\delta) \equiv {4 \over N} \sum_{k = 0}^{179} N_k \cos{2 p'_k},
\end{equation}
where $N_k$ is a number of galaxies with position angle $p'$.
The variance of $F$ is given by formula \cite{Pa97}:
\begin{equation}
\sigma^2(F)=8N^{-1}(179/180)^2.
\end{equation}
 
However, one should note that analyzed in \cite{Pa94} Function $F$ is a special
case of Hawley and Peebles \cite{h4} Fourier Test. The idea of this test
is following. If deviation from isotropy is a slowly varying function of
the angle $\theta$ (in our case $\theta=p'$) then
\begin{equation}
N_k = N_{0,k} (1+\Delta_{11} \cos{2 \theta_k} +\Delta_{21} \sin{2\theta_k})
\end{equation}
and we obtain the following expressions for the $\Delta_{i1}$ coefficients \cite{g2,g3, g10a,g11a}:
\begin{equation}
\Delta_{11} = {\sum_{k = 1}^n (N_k -N_{0,k})\cos{2 \theta_k} \over
\sum_{k = 1}^n N_{0,k} \cos^2{2 \theta_k}},
\end{equation}
\begin{equation}
\Delta_{21} = { \sum_{k = 1}^n (N_k-N_{0,k})\sin{2 \theta_k} \over
\sum_{k = 1}^n N_{0,k} \sin^2{2 \theta_k}}.
\end{equation}
where $N_k$ is the number of galaxies with orientations within the $k$-th  angular
bin, while $N_{0,k}$ denotes the expected number of galaxies in the $k$-th
bin. In the case of analysing of the distribution of the position angles 
all $N_{0,k}$ are equal $N_0$, which is also mean number of galaxies per bin.
It is easy to see that Function $F$ is in this case strictly related to 
$\Delta_{11}$  coefficient: $F=2\Delta_{11}$ (see \cite{g11a,g12a} for details).
 
 In our paper instead of equatorial coordinate system we used another,
supergalactic coordinate system (Flin   and God{\l}owski \cite{f4}) as a 
fundamental coordinate system, and the position of the coordinate system pole 
was varied (by $5^o$ at a time), both in supergalactic latitude $B$ and longitude $L$.

\section{The results}
 
Our main results were presented on the Figures 1-3. On the Figures 1 we 
presented  Function $F$ \cite{Pa94} for galaxies from Tully's NGC Catalogue 
depending on morphological types, while the results for the Figures 2 were 
restricted only for galaxies with certainly measured position angles. On every
maps we presented the value of analyzed function $F$ with respect to a 
chosen cluster pole, divided by its formal error $\sigma(F)$. In other words,
parameters $f=F/\sigma(F)$ were mapped. The cluster pole coordinates change 
along the entire celestial sphere. The resulting maps were analyzed for 
correlations of their maxima with important points on the maps. Maximum in 
$f$ function  mean that we have an excess of galaxies with position angles 
pointing in this direction. Minimum in $f$ function denotes a deficit of 
galaxies with position angles pointing in the particular direction. There is 
a clear interpretation for edge-on galaxies - minimum in position angles 
means an excess of normal to galaxy plane (believed as rotation axis of 
galaxies) pointing in the analyzed direction.
 
One can see from the Figure 1, that only for spiral galaxies (S) we obtained 
the $2\sigma$ extremum. Maximum is pointed near to the Local Supercluster plane in 
the longnitude $L$ of about $L=270^o$, while minimum is pointed to the 
direction perpendicular to the LSC plane. If we restricted our sample to the 
galaxies with certainly measured position angles (Figure 2) the result was 
similar but effect is much stronger, i.e. we observed $2\sigma$ extremum even
for all galaxies. Separately we analyzed a sample of 2227 UGC/ESO galaxies 
\cite{g2,g3} (Figure 3). The results for that sample were the same as for NGC
galaxies but positive extremum for all and spiral galaxies is on $3\sigma$ 
level.
 
Our results: - minimums of $f$ function in the direction perpendicular to the 
LSC plane for spiral galaxies (S) are in agreement with the results carried out 
by Aryal and Saurer \cite{a05}, who found weak preference of spin directions 
for the spiral galaxies belonging to the Local Supercluster. They found that 
the spin vector orientations of the spiral galaxies in the LSC tend to be 
oriented perpendicular to the LSC plane.

\section{Discussion and conclusions}
 
We analyzed the sample of galaxies belonging to the Local Supercluster
using proposed by Parnovsky, Karachentsev and Karachentseva Function $F$ \cite{Pa94}.
We found an excess of galactic position
angles pointing to the LSC plane in the direction about $270^o$ and
a deficit of galaxies with position angles perpendicular to LSC plane.
This effect is a weak one and connected with spiral galaxies. Our
results are in agreement with Aryal and Saurer \cite{a05} suggestion, 
who found  a weak preference of galactic spins for spiral galaxies, 
in the direction perpendicular to the LSC plane.
 
In our opinion the  function  proposed in \cite{Pa94} is useful but using full Peebles's
Fourier test (see for example \cite{h4,f4,g2,g3,g10,AR00,a05,Ar10,Ar11})
is more adequate because of exploring not only Peebles'  coefficient $\Delta_{11}$
but complete information connected with slowly varied function of angle $\theta$, 
$N_k = N_{0,k} (1+\Delta_{11} \cos{2 \theta_k} +\Delta_{21} \sin{2\theta_k} ......)$.
 
Possibility of existing large scale alignment, suggested by Parnovsky et al.
\cite{Pa94} needs more studies in future time. Recently some more papers 
implied such possibilities. Paz, Stasyszyn and Padilla  \cite{Paz11}
analyzing galaxies from the Sloan Digital Sky
Survey  Catalogue found that the  galaxy's  angular  momenta  are
aligned perpendicularly to the planes of large-scale structures,
while there is no such an effect for the small-scale structures. They
interpret this as consistent with their simulations based on the
mechanism of tidal interactions. Jones, van  der  Waygaert  and
Aragon-Calvo  \cite{Jones10} found that the spins of spiral galaxies
located within cosmic web filaments tend to be aligned along the
larger axis of the filament, which they interpret as "fossil"
evidence indicating that the action of large scale tidal torques
affected the alignments of galaxies located in cosmic filaments.

The evidence for the structural anisotropy of the small-large galaxy 
structures imside of the deep-field surveys likely for example the 
Jagellonian field \cite{Va1} was presented in \cite{Va2} where
the method of the effective ellipse of inertness was applied \cite{Va3}. 
The statistically significant anisotropy for the galaxy groups and clusters 
orientations  inside of the cone of the Jagellonian field R ~ 3000 Mpc is in 
the range 105-120 deg. \cite{Va2}. It also should be noticed that
Mandzos et al \cite{Man1,Man2}, during analyzis of UGC \cite{n1} (nord hemisphere)
and ESO \cite{l1}(south hemisphere) galaxies, found that galaxies  major
axes are distributed  anitropically.

Another possibilities of a large scale alignment were found in the
series of papers by Hutsemekers \cite{hu1,hu2,hu3} during analyzing of
the alignment of quasar polarization vectors.
In the paper Hutsemekers et al. \cite{hu1} based on a sample of
355 quasars with significant optical polarization  they found that
quasar polarization vectors are not randomly oriented over the sky.
The polarization vectors appear to be coherently oriented or aligned over
huge ($\sim$ 1 Gpc) regions at the sky. Furthermore, the mean
polarization angle $\theta$, appears to rotate with redshift at the
rate of $\sim 30^o$ per Gpc. While interpretations like a global
rotation of the Universe can potentially explain the effect, the
properties they observed qualitatively correspond to the dichroism
and birefringence predicted by photon-pseudoscalar oscillation within
a magnetic field. The origin of this results is still discussed
and possible interpretations were done for example by Hutsemekers
\cite{hu2,hu3,hu4} and recently by Agarwal, Kamal and Jain \cite{ag11}.


\begin{figure}
\vskip 20cm
\includegraphics{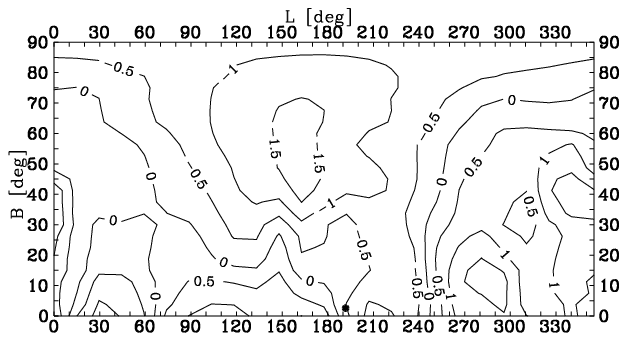}
\includegraphics{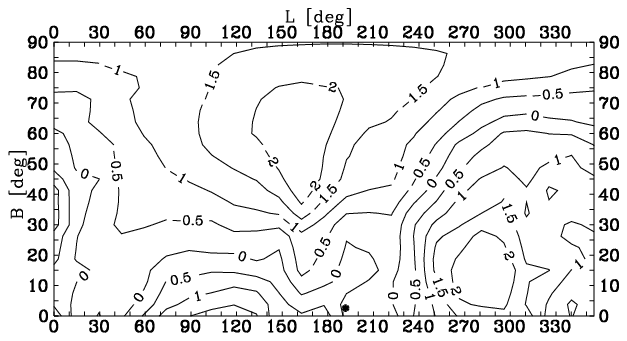}
\includegraphics{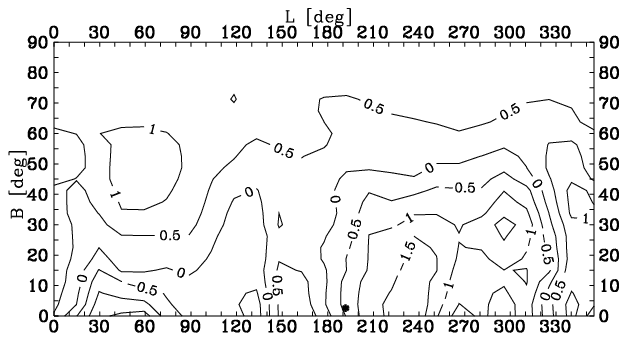}
\scriptsize
\caption{
Maps of
$f \equiv F/ \sigma(F)$ versus the chosen cluster pole in supergalactic
coordinates ($L$, $B$) for  NGC galaxies. The maps are presented for
ALL galaxies (upper panel), for S (middle panel) and NS (bottom panel)
sub-samples. On the map we indicated the important position,
the direction to the Local Supercluster centre (point).}
\end{figure}

\begin{figure}
\vskip 20cm
\includegraphics{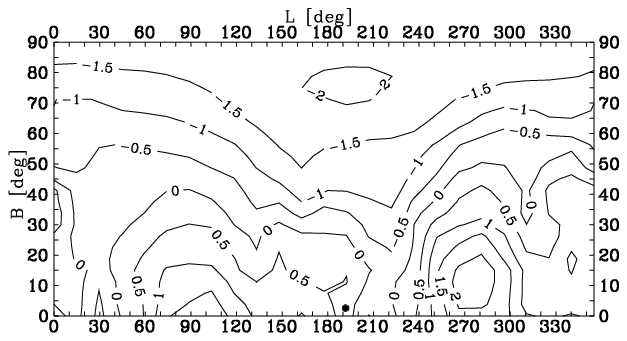}
\includegraphics{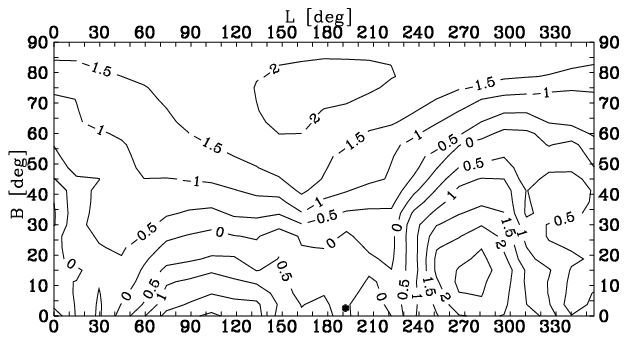}
\includegraphics{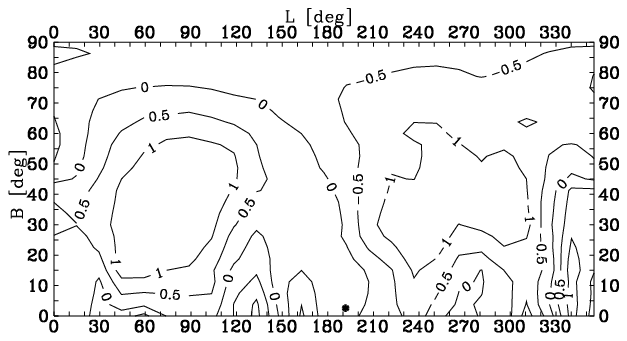}
\scriptsize
\caption{
Maps of
$f \equiv F/ \sigma(F)$ versus the chosen cluster pole in supergalactic
coordinates ($L$, $B$) for  NGC galaxies with certainly measured position
angles. The maps were presented for
ALL galaxies (upper panel), for S (middle panel) and NS (bottom panel)
sub-samples. On the map we indicated the important position,
the direction to the Local Supercluster centre (point).}
\end{figure}
 
\begin{figure}
\vskip 20cm
\includegraphics{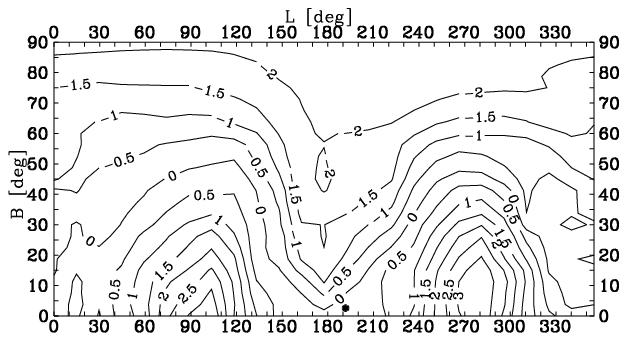}
\includegraphics{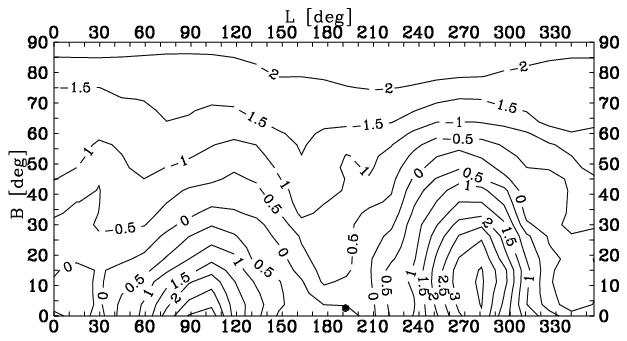}
\includegraphics{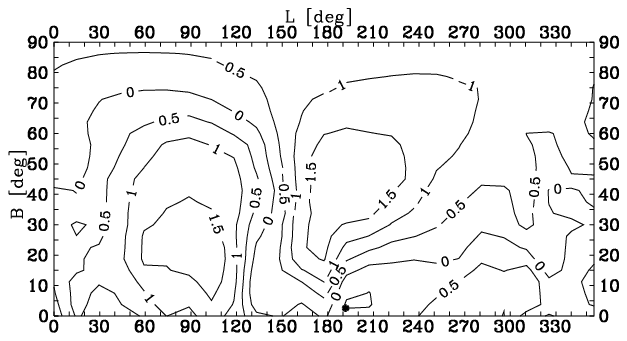}
\scriptsize
\caption{
Maps of
$f \equiv F/ \sigma(F)$ versus the chosen cluster pole in supergalactic
coordinates ($L$, $B$) for  UGC/ESO galaxies. The maps were presented for
ALL galaxies (upper panel), for S (middle panel) and NS (bottom panel)
sub-samples. On the map we indicated the important position,
the direction to the Local Supercluster centre (point).}
\end{figure}
 

\begin{thebibliography}{}
\bibitem {h4} Hawley, D. I., Peebles, P. J. E. 1975, Astron.J., 80, 477
\bibitem {Op70} {\"O}epik, E.J. 1970, Irish AJ, 9, 211
\bibitem {Ja78} Jaaniste, J., Saar, E. 1978, in: The large scale structures of the Universe., eds. M. S. Longair and J. Einasto, D. Reidel, Dordrecht (IAU Symp. 79), p.488
\bibitem {f4} Flin, P.,  God{\l}owski, W. 1986, MNRAS, 222, 525
\bibitem {g2} God{\l}owski, W. 1993, MNRAS, 265, 874
\bibitem {g3} God{\l}owski, W. 1994, MNRAS, 271, 19
\bibitem {g10} God{\l}owski, W., Flin, P., 2010,  ApJ  708,  902
\bibitem {AR00} Aryal, B., Saurer, W., 2000, A\&A, 364, L97
\bibitem {Pa94} Parnovsky, L. S., Karachentsev, I. D., Karachentseva, V., E. 1994, MNRAS, 268, 665
\bibitem {n1} Nilson, P. 1973, Uppsala General Catalogue of Galaxies, Astr. Obs. Ann. V, Vol.1: Uppsala
\bibitem {l1} Labuerts, A. 1982, ESO/UppsalaSurvey of the ESO B Atlas, ESO: Garching
\bibitem {k93}  Karachentsev, I. D., Karachentseva, V., E., Parnovsky, L. S. 1993, Astron. Nachr. 314, 97
\bibitem {Fl90} Fliche, H.H., Souriau, J.M., 1990 Astron. Astroph. 233, 317
\bibitem {f1} Flin, P., 1995 Comments Astrophys., 18, 81
\bibitem {Pa97} Parnovsky, L. S., Kudrya, Yu. N., Karachentsev, I. D. 1997, Astronomy Letters, 23, 504
\bibitem {t3} Tully, R. B. 1988, Nearby Galaxy Catalog, Cambridge
\bibitem {t2} Tully, R. B. 1987, APJ, 321, 280
\bibitem {n2} Nilson, P. 1974 Catalogue of Selected Non-UGC Galaxies, Uppsala Astr. Obs. Rep. 5: Uppsala
\bibitem {l2} Lauberts, A., Valentijn, E. 1989, The Surface Photometry Catalogue of the ESO-Uppsala Galaxies, ESO:Garching
\bibitem {g10a} God{\l}owski, W.,  Piwowarska, P., Panko, E., Flin, P., 2010, ApJ, 723, 985
\bibitem {g11a} God{\l}owski, W.,  2012, Astrophys. J., 747, 7 
\bibitem {g12a} Pajowska, P., God{\l}owski, W.,  Panko, E., Flin, P., 2012, arXiv 1301.5294 (2012, JPS 16 accepted)
\bibitem {a05} Aryal, B., Saurer, W., 2005a Astron. Astroph. 432, 431
\bibitem {Ar10} Aryal, B., Bachchan R.K., Saurer, W., 2010, Bull. Astr. Soc. India 38, 165
\bibitem {Ar11} Aryal, B., 2011, Research in Asronomy and Astrophysics 11, 293 (2010 arXiv 1010.5585)
\bibitem {Paz11} Paz, D. J., Sgr{\'o}, M. A., Merchan, M., Padill, N., 2011, MNRAS, 414, 2029
\bibitem {Jones10} Jones, B., van der Waygaert R., Aragon-Calvo M., 2010 MNRAS, 408, 897
\bibitem {Va1} Flin, P., Vavilova, I. B.  Proceedings of the 27th Meeting of the Polish Astronomical Society, Poznan, September 12-15, 1995, eds. by M. J. Sarna and P. B. Marks, 1996., p.63 
\bibitem {Va2} Gregul, A.Ya., Mandzhos, A.V., Vavilova, I.B. 1991, Astrophys. Space Sci., 185, 223
\bibitem {Va3} Vavilova, I. B.  1999, Kinematics Phys. Celest. Bodies, 15, No. 1,  69 
\bibitem {Man1}  Mandzos, A.V.,  Telnyuk-Adamchuk V.V.,  Gregul A.Ya.  1985, Sov.Astr.Lett. 11, 206 
\bibitem {Man2}  Mandzos, A.V., Gregul, A.Ya.  Izotova, I. Yu.   Telnyuk-Adamchuk V.V. 1987, AStrofisica, 26, 321,
\bibitem {hu1} Hutsemekers, D. 1998 Astron. Astrophys.  332, 410
\bibitem {hu2} Hutsemekers, D., Lamy, H. 2001 Astron. Astrophys. 367, 381
\bibitem {hu3} Hutsemekers, D., Cabanac, R.,  Lamy, H., Sluse,  D., 2005 Astron. Astrophys. 441, 915
\bibitem {hu4} Hutsemekers, D., Borguet, B., Sluse, D., Cabanac, R., Lamy, H. 2010
  Astron. Astrophys. 520, L7
\bibitem {ag11} Agarwal, N., Kamal, A., Jain, P., 2011 Phys. Rev. D. 83, 065014
 
 
\end{thebibliography}
\end{document}